\documentclass[9pt,twocolumn]{ieeetj}
\usepackage{cite}
\usepackage{amsmath,amssymb,amsfonts}
\usepackage{algorithmic}
\usepackage{graphicx,color}
\usepackage{textcomp}
\usepackage{xcolor}
\usepackage{hyperref}
\usepackage{booktabs}
\usepackage{float}  % Add this to the preamble

\hypersetup{hidelinks=true}
\usepackage{algorithm,algorithmic}
\def\BibTeX{{\rm B\kern-.05em{\sc i\kern-.025em b}\kern-.08em
    T\kern-.1667em\lower.7ex\hbox{E}\kern-.125emX}}
\AtBeginDocument{\definecolor{tmlcncolor}{cmyk}{0.93,0.59,0.15,0.02}\definecolor{NavyBlue}{RGB}{0,86,125}}
\def\OJlogo{\vspace{-4pt}$<$Society logo(s) and publication title will appear here.$>$}
\def\seclogo{\vspace{10pt}$<$Society logo(s) and publication title will appear here.$>$}
\def\authorrefmark#1{\ensuremath{^{\textbf{#1}}}}
\begin{document}
\receiveddate{XX Month, XXXX}
\reviseddate{XX Month, XXXX}
\accepteddate{XX Month, XXXX}
\publisheddate{XX Month, XXXX}
\currentdate{XX Month, XXXX}
\doiinfo{XXXX.2022.1234567}

\markboth{}{Author {et al.}}

\title{Securing Agentic AI: Threat Modeling and Risk Analysis for Network Monitoring Agentic AI System}

\author{Pallavi Zambare\authorrefmark{1},Member, IEEE, Venkata Nikhil Thanikella\authorrefmark{1},\\ and Ying Liu\authorrefmark{1}, Member, IEEE}
\affil{Department of Computer Science,Texas Tech University, Lubbock, Texas, USA}
 
\corresp{Corresponding author:Ying Liu (email:Y.Liu@ttu.edu).}
\authornote{ }

\begin{abstract}
When combining Large Language Models (LLMs) with autonomous agents, used in network monitoring and decision-making systems, this will create serious security issues. In this research, the MAESTRO framework consisting of the seven layers threat modeling architecture in the system was used to expose, evaluate, and eliminate vulnerabilities of agentic AI. The prototype agent system was constructed and implemented, using Python, LangChain, and telemetry in WebSockets, and deployed with inference, memory, parameter tuning, and anomaly detection modules. Two practical threat cases were confirmed as follows: (i) resource denial of service by traffic replay denial-of-service, and (ii) memory poisoning by tampering with the historical log file maintained by the agent. These situations resulted in measurable levels of performance degradation, i.e. telemetry updates were delayed, and computational loads were increased, as a result of poor system adaptations. It was suggested to use a multilayered defense-in-depth approach with memory isolation, validation of planners and anomaly response systems in real-time. These findings verify that MAESTRO is viable in operational threat mapping, prospective risk scoring, and the basis of the resilient system design. The authors bring attention to the importance of the enforcement of memory integrity, paying attention to the adaptation logic monitoring, and cross-layer communication protection that guarantee the agentic AI reliability in adversarial settings.
\end{abstract}


%\IEEEspecialpapernotice{(Invited Paper)}

\maketitle
\section{INTRODUCTION}
\subsection{OVERVIEW OF AGENTIC AI AND NEED OF AUTONOMOUS AGENTS IN NETWORK MONITORING}
 An agentic AI can be described as autonomous memory-based systems that engage in planning, relying, and communication with external tools in a behaviorally self-directed way. Compared to the reactive patterns of AI models that work on the basis of a single prediction, the agentic systems follow the perception-reasoning-acting loop. This allows them to make choices and aim at achieving objectives and changing with time. They are powered either by large language models (LLMs) or multimodal foundation models with the ability to reason in context, operate with long-term horizons, and co-ordinate multimodal actions that are not merely short-term responsive \cite{b1},\cite{b2}. Agents agentic architectures liberate agents stateful and compositional behaviors and actions, and planning beyond stimulus-response behavior, so are more suitable to dynamic environments where rule-based behavior cannot be adapted dynamically. An example is in network monitoring, where the conventional tools and systems such as static intrusion detection system (IDS) today as well as the threshold monitors are inadequate toward identifying new threats or zero-day threats \cite{b3}. The gap is closed in agentic AI by the ability to conduct on-the-fly telemetry analysis, context sensitive alerts, and auto-tunable parameters. A machine trained on an LLM could also make the difference between benign and malicious traffic influxes and respond, with hardly any guidance on the human side. Such agents can change constantly through feedback loops, and thus provide quick, adaptive solutions in cybersecurity situations. They improve resiliency through proactive mitigation of threats, cross-correlation of trends and detection gaps, thus the need to use agentic AI in the current dynamic threat environment \cite{b4}.
\subsection{GROWING DEPLOYMENT OF LLM-AUGMENTED AGENTS IN CYBERSECURITY}
 Large language models (LLMs) have allowed large scale deployments in cybersecurity to replace traditional, human-manned coordinating effort with semi-autonomous entities that perform threat, anomaly analysis, and response generation in real time. Based on the generative and contextual reasoning capabilities of LLMs, these agents can execute sophisticated work that typically needs a human background to conduct, including log parsing, incident summarization, and policy refinement. They are particularly suitable in the dynamic environment of such applications as network defense since they can be adapted to unstructured information and can learn through feedback. Some of the areas which LLMs are applicable include intrusion detection, phishing classification, alert triage, and vulnerability management. As an example, our model applied to a SIEM system using a transformer-based prediction correctly classified a substantial number of false positives with the help of correlating alert history with metadata context in the past \cite{b5}.
 \begin{figure}[h]
  \centering
  \includegraphics[width=8.5cm,height=5.05cm]{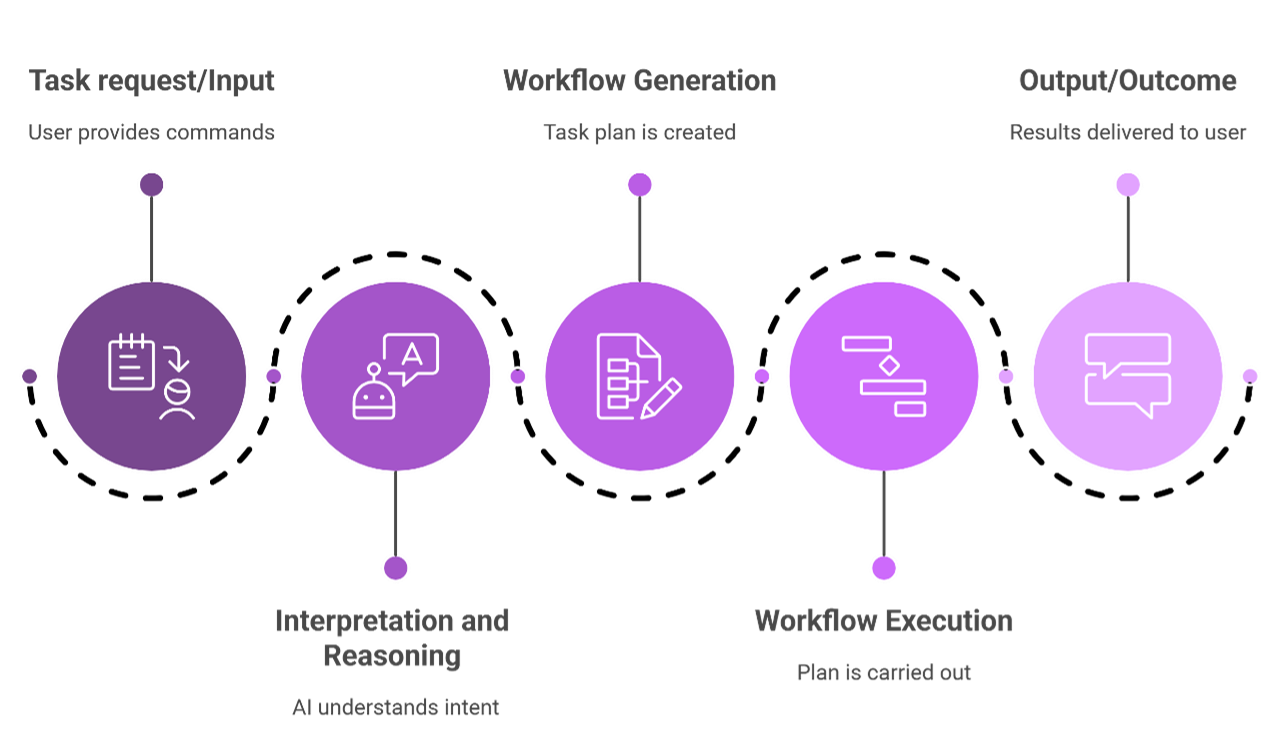}
   \caption{A generalized architecture of a single-agent Agentic AI system}
 %\Description{A woman and a girl in white dresses sit in an open car.}
  \label{Fig1}
\end{figure}
 Threat hunting as well as incident response also relies heavily on LLM-powered agents to help them extract indicators of compromise (IOCs) posted in live telemetry and execute automated responses in situations like firewalls updates \cite{b6}. These agents support workflow orchestration without scalable servers and distributed threat mitigation, and visibility and overall effectiveness in cloud-native infrastructures \cite{b7}. On the other hand, trust and control issues have been offered by such challenges as hallucinations, goal disparities, adversarial manipulation (e.g., memory poisoning), and partial observability. Their increase in decision making autonomy leads to an appetite among LLM agents to require more advanced threat modeling systems and architectural defenses in order to be deployed safely and in an auditable way.
\subsection{CHALLENGES: AUTONOMY, EXTERNAL TOOL USE, REASONING LOGIC — NEW THREAT CLASSES. EXISTING FRAMEWORKS (E.G., STRIDE, PASTA) ARE INADEQUATE.}
New security risks are presented by agentic AI systems, particularly large language models (LLMs) driven systems, predicated on an expansiveness that traditional rule-based or supervised AI could not support. Such agents have advanced features including autonomous reasoning, persistent memory, adaptive planning, and external tools invocation, enlarging their scale of functionality considerably, and, respectively, the surface of attack. The threats arising include goal misalignment, memory poisoning, and multi-stage reasoning hijacks because of the changing behavior of agents in the dynamic environment. Such frameworks as STRIDE and PASTA do not have the ability to deal with such complex interactions, as all of the architectures are modeled as static, and are not deep in modeling semantic or emergent behavior. Overcoming these limitations, this work implements the MAESTRO framework that divides the architecture of an agent into 7 interrelated layers so that the threat localization at different layers and the risk scoring based on a specific AI are possible. The questions the research produces respond to are the following: 
\begin{enumerate}
    \item How do threats emerge in autonomous agents?
    \item  How can they be mapped and evaluated efficiently using MAESTRO ?
    \item What mitigation strategies are the best?
\end{enumerate}
It gives the contextual MAESTRO mapping, threat taxonomy, a risk scoring model based on severity, and a real-world validation in the scenario of a deployed LLM-based agent in a live telemetry monitoring attitude.
\section{Background and Related Work}
\subsection{Existing Security Frameworks}
The established models of security, such as STRIDE, PASTA, OWASP Top 10 have long been utilized in secure software development and insufficiently accommodate the agentic AI systems that feature autonomy and dynamic reasoning. To show how to apply STRIDE to structured systems such as infotainment, Das et al. \cite{b8} found 34 threats. Tete et al. \cite{b9} and Gulen et al.\cite{b10} have also generalized STRIDE to LLM-based applications and have typified concerns such as injection and privilege abuse. Nevertheless, STRIDE does not model emergent behavior, cognitive reasoning of AI agents very well. Juuso \cite{b11} pointed out that the threats associated with invoking a tool and the cognitive subsystems require the scope of STRIDE to be extended or supplemented.

PASTA is an attacker-oriented, process-focused approach targeted at modeling the behavior of threat through the cycle of a system. Even though theoretically inclusive, its boundaries of the system are relatively fixed, which is why it might not be fully applicable to those environments that pursue agentic AI. Juuso \cite{b11}observes that though PASTA provides detailed threat flow, it is not flexible to systems that have a learning and adaptive behavior during runtime. It assumes attacker goals are pre-determined and data flows are fixed, which disrupts its applications in black-box systems that commonly cause new vulnerabilities due to unpredictable modifications of their state. The OWASP Top 10 on LLM Applications captures the widening gap between the conventional application security and the LLM-related threats. Vulnerabilities it points are over-agency, hallucination, prompt injection, and insecure memory. Tete et al.\cite{b9} depicts these weaknesses of chatbot agents, and Dev et al. \cite{b12} use the OWASP recommendations to construct live-time security checklists within the Guard Rail system. Such OWASP categories as unsafe functions calls and overreliance, which are of particular interest to agents that work with APIs or external code, are highlighted by Khan et al. \cite{b13}.

Although OWASP is related to emergent LLM behaviors, it does not have structural modeling of reasoning and planning levels. Classical theories such as STRIDE and PASTA are unable to deal with agentic modularity- memory systems, perception layers and planning engines. To address this gap, Huang et al. \cite{b14} with an introduction to model threats at multiple levels (e.g., at layers of Input, Memory, Reasoning, Planning, and Action Execution), providing a new framework referred to as MAESTRO. Dev et al. \cite{b12} promote the conversational and systematic controls like hallucination filters and safe delegation. Alwaheidi \cite{b15} emphasizes the necessity of threat monitoring that should be performed continuously, and Mollaeefar et al. \cite{b16} describe such dangers as memory poisoning, failures in recursive planning, which are not considered by conventional models.

\subsection{RELATED WORK}
 The articles conducted recently highlighted the problem of security risks posed by systems that use LLM in operational settings. Krishnamurthy \cite{b17} explored the application of generative AI in cyber defense and realized that hallucinations and coincidental overreliance might lead to false detections of threats in operational networks. Coletta et al. \cite{b18} considered LLMs finding computable code and alerted to vulnerability through recursive generation and not-quite-secure chain of APIs. The above problems led Clement et al. \cite{b19} to suggest an end-to-end threat modeling approach to LLMs that combines STRIDE-DREAD with injection simulations that expose the programming to complex attack surfaces in autonomous pipelines. These findings were later confirmed by Gulen et al. \cite{b10} regarding AI as a service systems with the relevant risk being the possibility of unauthorized access and manipulation of conclusions since it is modular, service-based deployments.
 
Structural vulnerabilities in agentic systems have been dealt too. Messaad \cite{b20} proposed a modular model founded on feedbacks, interface indeterminacy, and poisoning of memory. Selin \cite{b11} emphasized that the older models of threat and present-day agent frameworks are incompatible. Chen et al. \cite{b21} stressed that tools summoned by LLMs such as shell commands or APIs are very vulnerable to indirect prompt injection, especially in unsupervised execution chains. Another research done showed that the MAESTRO framework was highly effective to identify memory abuse, planning misalignment, and unsafe formations of tools in agentic deployments (Talla et al. \cite{b22}).

Self-delegation and hallucinated planning are emergent behaviors that also came to the attention. Huang \cite{b23} outlined the risks of coordination in the collective of agent systems that share memory or APIs whose trust boundaries are not defined. GitHub and Github Classroom \cite{b10} were also compromised at inference times by adversarial orchestrated API calls in AIaaS contexts. Sere et al. \cite{b24} reported the abuse of the tools when they were taken out of policy as a consequence of low validation. Fuchs et al. \cite{b25} provided information about fails to do blind delegation in multi-agent planning. Zhang et al. \cite{b26} suggested a policy engine in run time that interrupts a mismatch between planning and actions to avoid any damage that can be caused by a hallucinated behavior.

\subsection{LITERATURE SUMMARY AND GAP}
Although these frameworks, namely STRIDE, PASTA and OWASP, are very helpful, they do not cover the reasoning stratum and autonomous operations of agentic AI systems, bounding up on defined attack surfaces. Their incapacity to follow the path of threats in terms of memory, planning, and execution of tools has been already acknowledged in the literature. In spite of dealing with secluded vulnerabilities, OWASP LLM Top 10 lacks structural granularity and cannot model emergent behaviors, which take place in multi-agent systems. Conversely, this paper implements and applies the MAESTRO framework in order to overcome such drawbacks within a threat modeling framework characterized by a layered and, agent-centric perspective. This work is also practically more solid than the other literature that is driven by the paradigm or still remains theoretical. It plays a part in generating the domain-specific threat list, establishing a quantitative risk-ranking program, and authenticating the mitigation strategies within an as-soon-as-you-see-it (as-soon-as-you-collaborate-on-it) telemetry capability processing environment. In this way, the work shifts into a practical presentation after performing the conceptual discussion and underlies the practicability of MAESTRO as a future real cybersecurity defense against agentic systems. 

\section{AGENT ARCHITECTURE AND THREAT LANDSCAPE}
\subsection{ARCHITECTURE OVERVIEW}
The suggested system is an autonomous agent based on the Large Language Model (LLM) and used to monitor the network in real time, having the potential to reason, detect, and react to unusual activities. This agent is released in a framework of a company or cloud-platform where it can monitor the telemetry based on the packets-level, evaluate the performance indicators, identify a threat in terms of security and provide an actionable representation, which can  be created autonomously with integrated memory and reasoning logic. The Figure \ref{DM1} below shows the architecture of the network monitoring agent which is based on the LLM, representing the data flow and interaction between the main components.
\begin{figure}[h]
  \centering
  \includegraphics[width=2.27cm,height=4.13cm]{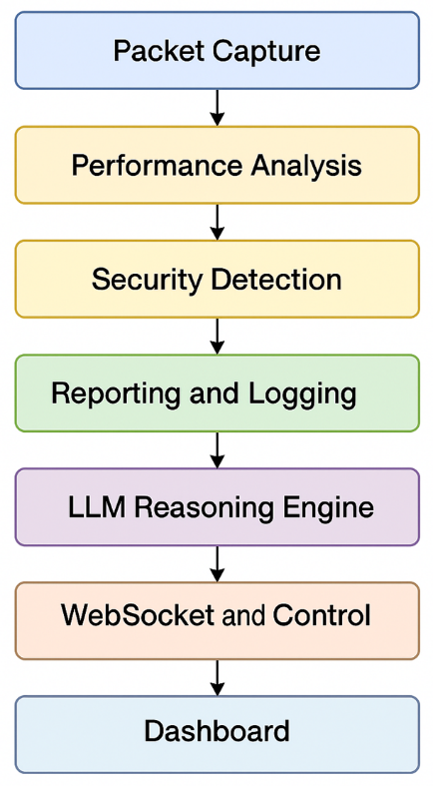}
   \caption{Architecture of the LLM-based network monitoring agent, depicting the data flow and interaction among core modules.}
 %\Description{A woman and a girl in white dresses sit in an open car.}
  \label{DM1}
\end{figure}
This architecture is divided into seven modular components which each perform one specific action to achieve the situational awareness and measure of responsiveness of the entire agent:\\ \textbf{Packet Capture Module:}It is a module that captures the real time network traffic through the various network interfaces, by utilizing the libpcap or similar network sniffing applications. It allows extensive examination of the packets as they are being sent in or sent out, this would provide raw material as security as well as performance analysis.\\
\textbf{Performance Analysis Module:} This layer keeps track of latency, jitter, bandwidth usage and other QoS values. It is time-series based and gives a picture of system health and points out regressions in performance that can signal configuration anomalies or resource bottlenecks.\\
\textbf{Security Detection Module:} Based on the use of pattern recognition, behavior analysis, and learned signature database, this module will detect any potential threats of a type of DoS attack, port scanning, or credential brute-forcing. It is used along with LLM-based reasoning to do context-sensitive correlation.\\
\textbf{Reporting and Logging Module:}This module compiles processed data in structured logs and reports. It also helps in periodical archiving, anomaly logging, in order to carry out a forensic investigation.\\
\textbf{LLM Reasoning Engine:}
The heart of the agent is a sharply optimized transformer-based model with the abilities to perform contextual examination, inference shops by use of memories, and independent response creation. The engine wires the system to read logs, correlate statistics and summarize events, plan remediation actions with reference to historical events and real-time indicigators.\\
\textbf{WebSocket and Control Layer:}Enables bidirectional interactions between the system and remote dashboards as well as control clients. This layer makes certain that the agent is able to take upgrades, send up alerts, and to synchronize courses of action within the monitoring ecosystem.\\
\textbf{Interactive Dashboard: }It is an interface which sits at the user level, displaying the latest trends of traffic, summary of alerts, system health status and an audit trail of actions taken by the agent. This module helps administrators to read, and override or accept automatically made decisions by the agent.\\
Combined, these modules will form a closed-loop monitoring system where the agent does not only identify and report about a threat but also models over telemetry to provide proactive actions or issue high-confidence alerts. The system architecture has the capability of being both reactive and predictive thereby having a robust nature to the dynamic nature of threat landscapes.
\subsection{EXPANDED ATTACK SURFACE}
The agentic AI is more complex architecture-wise creating a wider and deeper threat surface. In sharp contrast to conventional systems, autonomous agents are vulnerable to dynamic goals, memory loss, tool invocations and long-temporal-horizon reasoning. Such systems are vulnerable to both overt attacks of input or output, as well as of internal logic, memory of the historical past, planning capabilities, and interface to external tools. Here we classify ten imperative classes of threats that occur in the LLM-based network surveillance agents. Every threat is also defined by a given point of exploitation-e.g. instruction manipulation, multi agent poisoning- and matched to corresponding system layers by the means of the MAESTRO framework. The aim is to explain how the behaviour driven by the reasoning, memory-reliance, and unreliant to an outside source of autonomy expands the possible market of attack, and necessitates a more sophisticated modeling method than the common AI mechanisms with regard to security. As this Table 1 shows, the above threats have brief definitions, real-life application, and associations with MAESTRO layers.
\begin{table*}[htbp]
\caption{Threat Summary }
\centering
\small
\setlength{\tabcolsep}{4pt}
\renewcommand{\arraystretch}{1.7}
\begin{tabular}{@{}p{3cm}p{4.2cm}p{4.2cm}p{3.5cm}@{}}
\toprule
\textbf{Threat Name} & \textbf{Definition (Brief)} & \textbf{Example Use Case} & \textbf{MAESTRO Layer(s)} \\
\midrule
Instruction Manipulation & Alters input prompts to redirect agent behavior & Injected log entry modifies anomaly thresholding & Perception + Action \\
\midrule
Goal Manipulation & Shifts agent’s objectives via ambiguous feedback & Agent favors performance over security after drift & Planning + Reasoning \\
\midrule
Chain-of-Thought Manipulation & Corrupts stepwise inference of the LLM & Agent interprets attack as normal backup & Reasoning + Planning \\
\midrule
Memory \& Context Manipulation & Poisons historical memory/context for decisions & Altered alert history causes missed repeat attack & Memory + Reasoning \\
\midrule
Critical System Interaction & Misuses tool invocation to harm system & Malicious firewall update via spoofed trigger & Action \\
\midrule
Planning \& Reasoning Exploit & Degrades long-term decisions through input shaping & Attacker simulates load balancing to hide DDoS & Planning + Memory \\
\midrule
Resource Exhaustion & Overloads agent with data to reduce functionality & Flooding logs exhaust token capacity & Infrastructure \\
\midrule
Knowledge Base Poisoning & Corrupts learned information or references & False threat Intel causes misclassification & Knowledge + Reasoning \\
\midrule
Supply Chain Compromise & Inserts vulnerability via plugins or models & Tainted plugin leaks sensitive data & Infrastructure \\
\midrule
Multi-Agent Exploitation & One agent sabotages others via shared memory & Poisoned memory misguides downstream classification & Memory + Coordination \\
\bottomrule
\end{tabular}
\label{tab:agent_threats_maestro}
\end{table*}

\section{THREAT MODELING USING MAESTRO}
\subsection{OVERVIEW OF MAESTRO FRAMEWORK}
The MAESTRO framework introduces a multilayered threat modeling framework targeted to deal with security challenges in the context of the emergent agentic AI systems. In contrast to the traditional monolithic models, the operational stack of an agent is broken down in MAESTRO into seven connected layers, enabling a specific localization of the vulnerabilities and the possibility to employ quick mitigation strategies that involves only one layer. L1, Foundation Models, contains pre-trained and fine tuned LLM which do basic reasoning. Data Operations in Layer 2 provides data pipelines, labeling and storage. The orchestration and decision-making between agents is in the Layer 3, Agent Frameworks. Layer 4, Deployment and Infrastructure, includes topics on containerized hostings and cloud-based hosting. Layer 5, Evaluation and Observability, performs monitoring and integrity of the systems and Layer 6, Security and Compliance, tackles privacy, access control and regulatory assurance. The highest layer, Layer 7 Agent Ecosystem, facilitates the collaboration and interaction of multi-agent with external items. This model fits properly to the network agent which is suggested by the research in the context of LLM. The reasoning logic is in Layer 1, real time telemetry handling in Layer 2 and distributed alerting in Layer 7. This stratified breakdown allows complex (and emergent) and threats to be detected and also enables contextualized mapping of the behavior, autonomy, memory access and tool invocation of agents, a set of capabilities that could not be canvased by linear threat models of past.
\begin{figure}[h]
  \centering
  \includegraphics[width=9cm,height=4.5cm]{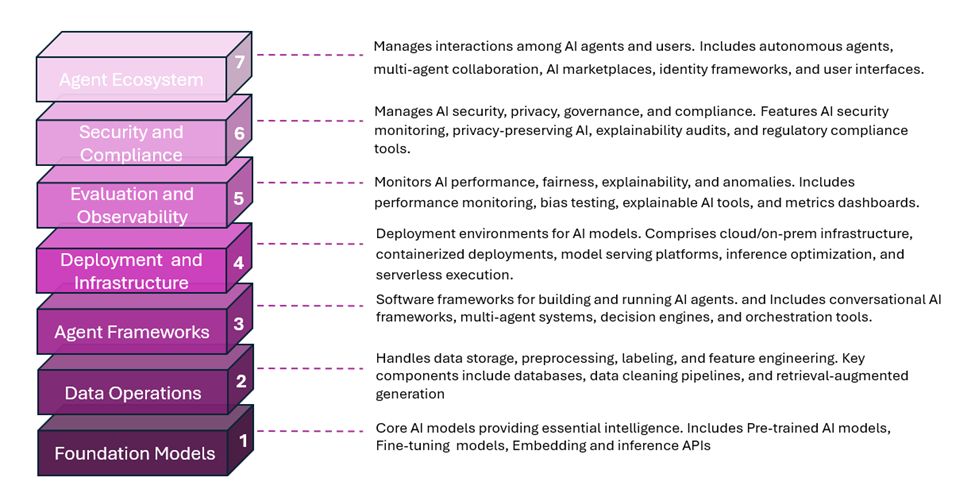}
   \caption{Complete seven-layer MAESTRO architecture}
 %\Description{A woman and a girl in white dresses sit in an open car.}
  \label{fig3}
\end{figure}
Classification of threats through holistic threat analysis and implant detection systems In our system of network monitoring agents, which combine telemetry ingestion, performance diagnostics, and LLM-based reasoning to achieve real-time detection of anomalies, MAESTRO can offer a convenient basis of comprehensive threat analysis and implant detection. Our architecture(Figure 3) is described as follows by the seven layers :\\
\textbf{L1: Foundation Models:}The LLM that carries out fundamental inference on pattern of traffic and anomalies in performance.\\
 \textbf{L2: Data Operations:}Data pipelines that aggregate, filter, and model network performance statistics.\\
\textbf{L3 Agent Frameworks:}Agent Level 3: Pre-planning and execution logic that calls out modules of detection and analysis on basis of system states.\\
\textbf{L4: Deployment \& Infrastructure: }FastAPI back-end, WebSocket APIs and containerized microservices that constitute the run-time environment.\\
\textbf{L5: Evaluation \& Observability:}The performance logs, anomaly detection metrics, and dashboards that assist the operator to situational awareness.\\
\textbf{L6: Security and compliance: }Authentications protocols, API level security controls and auditing mechanisms to reason and make decisions.\\
\textbf{L7: Sensor Ecosystem: }Agent ecosystem Interfaces to human operators and other external agents that might coordinate with, or communicate with, the monitoring system.\\
Such mapping serves to help MAESTRO, not only as a theoretical framework, but also as a method of operation, based on the design of our autonomous monitoring agent in the real world. The classification of threats based on it, the scoring of the magnitude of their impact and the design of mitigation strategies in the following sections are based on this structure.
\begin{table*}[htbp]
\caption{Threat to Layer Mapping}
\centering
\small
\setlength{\tabcolsep}{5pt}
\renewcommand{\arraystretch}{1.2}
\begin{tabular}{@{}p{3.5cm}p{3.2cm}p{2.8cm}p{5.5cm}@{}}
\toprule
\textbf{Threat} & \textbf{Primary Layer} & \textbf{Cross-Layer Impact} & \textbf{Brief Description} \\
\midrule
1. Instruction Manipulation & L7 – Agent Ecosystem & L1, L3 & Alteration of operator commands or prompts to guide agent behavior maliciously. \\
\midrule
2. Goal Manipulation (Agent Drift) & L3 – Agent Frameworks & L1, L6 & Subtle alteration of long-term goals due to poisoned reasoning chains or telemetry drift. \\
\midrule
3. Chain-of-Thought Manipulation & L1 – Foundation Models & L3 & Prompt injection or bias influencing internal stepwise reasoning of the agent. \\
\midrule
4. Memory and Context Manipulation & L1 – Foundation Models & L2, L3, L5 & Injection or corruption of episodic memory influencing future responses or analysis. \\
\midrule
5. Critical System Interaction & L4 – Deployment \& Infrastructure & L6, L7 & Unauthorized or misaligned tool use leading to destructive API actions. \\
\midrule
6. Planning and Reasoning Exploitation & L3 – Agent Frameworks & L1, L5 & Triggering incorrect logic flows or plan execution via crafted observations. \\
\midrule
7. Resource Exhaustion & L4 – Deployment \& Infrastructure & L2, L5 & Overloading agent resources (CPU, memory, tokens) to cause denial-of-service or degraded performance. \\
\midrule
8. Knowledge Base Poisoning & L2 – Data Operations & L1, L3 & Injection of malicious or misleading data into retrieval-augmented pipelines or local KBs. \\
\midrule
9. Supply Chain Compromise & L6 – Security \& Compliance & L1–L4 & Backdoored models, libraries, or components compromising the agent before deployment. \\
\midrule
10. Multi-Agent Exploitation & L7 – Agent Ecosystem & L1–L6 & Exploiting coordination protocols among agents to induce misalignment or cascading failures. \\
\bottomrule
\end{tabular}
\label{tab:threat_layer_mapping}
\end{table*}
\subsection{THREAT-TO-LAYER MAPPING}
Our system of network monitoring represents an agentic behavior which adds a magnifying window of vulnerability towards an increased threat surface especially through incorporation of LLMs, dynamic planning, tool invocation, and contextual memory. In contrast to traditional AI systems, such threats most commonly arise because of a reasoning path, modify the memory state or because of some form of interaction with the external system, occasionally spreading to more than one operational level. MAESTRO framework allows making systemic mapping of these threats to respective system layers, so that the security assessment would be localized and would be cross-referenced. Each threat is initially paired with the primary layer in which it is present or most serious and then paired with essentially all other layers it could propagate to or interact with in downstream process.Table \ref{tab:threat_layer_mapping} below shows the threat to layer mapping.
\subsection{RISK SCORING METHODOLOGY}
The autonomous and dynamic API technology design poses a multidimensional threat Dominion, with risk vectors varying drastically with respect to severity, frequency, and overall exploitability. Such assessment and prioritization of threats should therefore be conducted in a systematic way. Prioritising risks is a way of making the security stakeholders focus their mitigation strategies on the threats with maximum potential of causing harm to system stability and integrity. To ease this, the current research will use qualitative threat scoring model where critical risk score (R) of each threat is calculated as:
\begin{equation}
R = P \times I \times E
\end{equation}
\noindent
Where:
\begin{itemize}
    \item \textbf{P (Likelihood)} represents the approximation of the probability of the threat occurring under operational or adversarial conditions;
    \item \textbf{I (Impact)} refers to the level of consequences that are likely to occur in case of the accomplishment of the threat;
    \item \textbf{E (Exploitability)} is a measurement of the susceptibility of the threat being executed or guided by an adversary, considering system exposure, access requirements, and attack complexity.
\end{itemize}
All these three dimensions are checked with the help of an ordinal scale on the basis of qualitative evaluations:
\begin{table}[H]
\caption{Qualitative Assessment to Ordinal Value Mapping}
\centering
\small
\setlength{\tabcolsep}{6pt}
\renewcommand{\arraystretch}{1.2}
\begin{tabular}{@{}p{4.5cm}c@{}}
\toprule
\textbf{Qualitative Assessment} & \textbf{Ordinal Value} \\
\midrule
Low & 1 \\
Medium & 2 \\
High & 3 \\
\bottomrule
\end{tabular}
\label{tab:qualitative_mapping}
\end{table}
This model reflects the layered and interconnected nature of agentic systems, where minor vulnerabilities—when combined with autonomous planning, long-term memory, or tool invocation—can lead to significant systemic disruptions. By quantifying risks along these dimensions, the framework supports rigorous threat comparison, grounded in both technical feasibility and operational consequence.
\subsubsection{Illustrative Risk Score Examples}
In order to illustrate the interpretability of the model, one can look at the following scenarios:
\begin{itemize}
    \item The probability of occurrence, impact, and exploitability are low and high respectively ($P=1$, $I=1$, $E=3$), which makes the risk score $R = 1 \times 1 \times 3 = 3$ with low backward effect.
    \item A balanced threat in all dimensions ($P=2$, $I=2$, $E=2$) has a risk value of $R = 2 \times 2 \times 2 = 8$, implying moderate consideration.
    \item A hard-to-exploit, high-impact, and high-likelihood threat ($P=3$, $I=3$, $E=1$) yields a risk rating of $R = 3 \times 3 \times 1 = 9$, which indicates a high priority area.
\end{itemize}
\subsubsection{application to Identified Threats}
With the help of this model, Section 3.2 identified ten threats, which were considered separately. The probability, effect, and exploitability indicators were rated depending on the ability of each threat to interfere with the critical agentic capabilities of the action, either by manipulating memory, interrupting reasoning, or gaining unauthorized access to tools. Further contextualization of these assessments was based on analysis of the primary propagation and cross-layer propagation of each threat in the MAESTRO framework in such a way that threats with overall high architectural impact would be correctly placed high in priority. The findings are aggregated in form of a threat-risk matrix, which includes:
\begin{itemize}
    \item  Qualitative evaluation of likeliness, effect, and usability,
    
    \item  The Threat Risk score (TRS) computed to prioritize
    
    \item  Association with particular MAESTRO layers that were mapped.
\end{itemize}
The aggregated threat landscape can help to prioritize the threats in a well-organized and evidence-based manner to mitigate risks, which is the purpose of this matrix as shown in Table 4 below. Table 4 below shows the MAESTRO-Based Threat Risk Matrix for Network Monitoring Agent.
\begin{table*}[htbp]
\caption{MAESTRO-Based Threat Risk Matrix for Network Monitoring Agent}
\centering
\small
\setlength{\tabcolsep}{3pt}
\renewcommand{\arraystretch}{0.5}
\begin{tabular}{@{}p{3.2cm}p{3.2cm}p{3.3cm}p{1.8cm}p{1.5cm}p{2cm}p{1.5cm}@{}}
\toprule
\textbf{Threat} & \textbf{Primary Layer (MAESTRO)} & \textbf{Cross-layer Impact} & \textbf{Likelihood (P)} & \textbf{Impact (I)} & \textbf{Exploitability (E)} & \textbf{Risk Score} \\
\midrule
1. Input-Induced Behavior Manipulation & Agent Frameworks (L3) & Data Operations (L2), Evaluation \& Observability (L5) & High (3) & Medium (2) & Medium (2) & 12 \\
\midrule
2. Goal Manipulation & Agent Frameworks (L3) & Foundation Models (L1), Data Operations (L2) & High (3) & High (3) & Low (1) & 9 \\
\midrule
3. Chain-of-Thought Manipulation & Foundation Models (L1) & Data Operations (L2), Evaluation \& Observability (L5) & High (3) & High (3) & Medium (2) & 18 \\
\midrule
4. Memory \& Context Manipulation & Agent Frameworks (L3) & Data Operations (L2), Evaluation \& Observability (L5) & Medium (2) & High (3) & Medium (2) & 12 \\
\midrule
5. Critical System Interaction & Deployment \& Infrastructure (L4) & Agent Ecosystem (L7), Agent Frameworks (L3) & Medium (2) & High (3) & Medium (2) & 12 \\
\midrule
6. Planning \& Reasoning Exploitation & Agent Frameworks (L3) & Evaluation \& Observability (L5), Security (L6) & High (3) & High (3) & Medium (2) & 18 \\
\midrule
7. Resource Exhaustion & Deployment \& Infrastructure (L4) & Agent Ecosystem (L7), Evaluation \& Observability (L5) & High (3) & High (3) & High (3) & 27 \\
\midrule
8. Knowledge Base Poisoning & Data Operations (L2) & Foundation Models (L1), Agent Frameworks (L3) & High (3) & High (3) & Low (1) & 9 \\
\midrule
9. Supply Chain Compromise & Data Operations (L2) & Deployment \& Infrastructure (L4), Foundation Models (L1) & Medium (2) & High (3) & Medium (2) & 12 \\
\midrule
10. Multi-Agent Exploitation & Agent Frameworks (L3) & Agent Ecosystem (L7), Deployment \& Infrastructure (L4) & Medium (2) & High (3) & High (3) & 18 \\
\bottomrule
\end{tabular}
\label{tab:risk_scoring}
\end{table*}
\section{MITIGATION STRATEGIES AND DESIGN RECOMMENDATIONS}
\subsection{PREVENTION}
The issue of reducing the security risks in agentic AI systems depends on a multilayered approach through the use of effective prevention measures. An initial mitigation step is input validation and clean-up which means that missing or malicious instructions cannot affect the goal planning in the agent or memory contexts. Multiple works underline how it is possible to prevent timely injection and manipulation of chain-of-thought arguments with help of restricting untrusted user inputs via pre-processing and content sanitization \cite{b27}. LLM guard-rails content filter, quick limit, as well as safety-alignment modules, can serve as first-line protections in the masking noxious productions or actions. Such alarms are especially essential in situations where LLMs work with open-ended and multi-turn settings \cite{b28}. Also, by sandboxing agents and restricting them to access a certain set of tools by employing wrapper based interface sandboxing agents can not access any of the sub system which is not within the control of said agents \cite{b29}.
One such commonly recommended approach is to have capability-based access control, limiting what an agent may do, instead of limiting it based on who it is. This is also enforced in multi-agent contexts, in which otherwise, delegation and sharing of memory context can result in privilege escalation \cite{b30}. In addition, the paper suggests adding layer-based controls like L2 memory isolation and L3 planner verification which are noted as the best practices of reducing lateral vulnerabilities in the context of layered deployments \cite{b31}. The other important prevention measure is the reward shaping and constraint learning that can provide training in avoiding unsafe or sub-optimal decision trajectories \cite{b32}. This is necessary in safe autonomy when agents are functional with minimal input of the human agent. Lastly, a zero-trust architecture approach in which each action is authenticated, each resource secured, and each interaction tracked are proposed to be applied in agentic systems \cite{b33}. This goes very much in line with the multi-layers perspective of security incorporated in the design of MAESTRO framework that supports the importance of embedded verification mechanisms across the agent lifecycle \cite{b34}.
\begin{figure}[h]
  \centering
  \includegraphics[width=8cm,height=3.97cm]{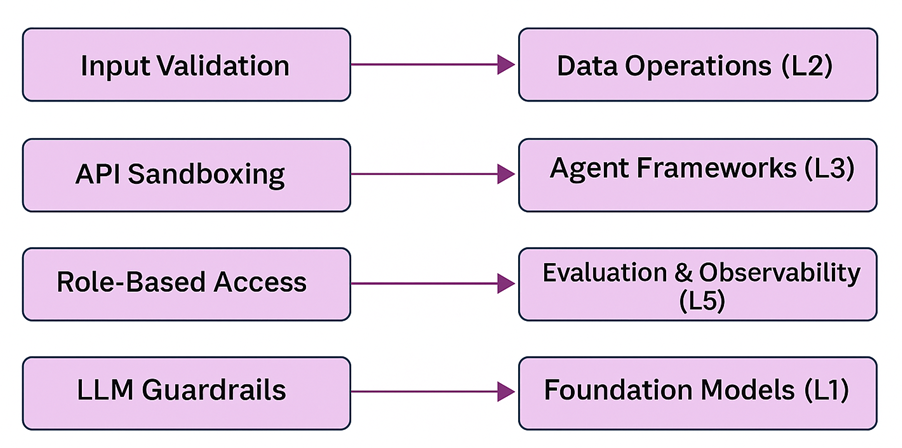}
   \caption{Prevention Strategies for securement of an autonomous agent using MAESTRO Layer}
 %\Description{A woman and a girl in white dresses sit in an open car.}
  \label{fig4}
\end{figure}
\subsection{DETECTION AND RESPONSE}
Although preventive measures are becoming more significant in terms of preventing malicious work in advance, another very crucial shield in the autonomous systems would be a non-negligible ability to detect and respond to malicious activities. Since the agentic AI behavior is unpredictable, adaptive and non-static, it is not realistic to expect that all the attack vectors can be eliminated in advance. Hence, runtime monitoring and reactive security layer is important to sustain integrity and traceability of operations. Real-time anomaly detection is one of the elements of this layer. Due to the ongoing telemetry data collection process that incorporates traffic patterns, agent decision latencies, resource consumption, and deviations to expected operating parameters, these agents can indicate signs of compromise or drift against anticipated operating parameters. Such process usually makes use of threshold checks, statistical modeling of agents that may shift due to a malicious agent or a more elaborate ML-based anomaly detection based on the outputs of the reasoning of the agent or on the execution path.

During identification of trouble, rollbock mechanisms are very essential in maintaining stability of systems. This can comprise going back to already understood safe policy checkpoints, returning LLM contexts, or turning off modules to stop the spread of the anomaly. Such mechanisms prevent that temporary corruption, or unintended or possibly malicious steps in the reasoning process can persist or propagate. Also, there is the factor of forensic logging, which is essential both during post-incident analysis and compliance. All interactions be it inputs, intermediate reasoning chains, API calls, and final output should be stored in tamper-proof so as to achieve traceability. With these logs it is possible to recreate attack chains, identify vulnerabilities being exploited, and produce threat intelligence that is fed back to lead to more effective heuristics. A combination of these abilities such as anomaly detection, rollback, and forensic logging becomes the main construction of the response approach in agentic AI landscapes. They allow the system to not just respond to threats quickly but also to learn after being attacked thereby bringing iterative hardening of agent behavior in long term.
\subsection{DEFENSE-IN-DEPTH}
A powerful agentic AI system should not use one layer of security. Since the architecture of such products is multi-component, layered, the strategy should be defense-in-depth which should aim at having security not undermined even when individual layers are breached. It is a method that shares security throughout the stack, which correlates well with the MAESTRO frameworks and run-time actions of the system. All the layers of the MAESTRO architecture(Figure 4)play a role in the security:\\
\textbf{L1 Foundation models: }Foundation models have guardrails and fine-tuning constraints to avoid undesirable outputs in inference. Prompt and response filters create filtered bad queries so that malicious queries are filtered out before creating unsafe chains of reasoning.\\
\textbf{L2 Data Operations:}Important telemetry and contextual data are secured through memory isolation, with policies set on data sanitation, and access. This blocks usage sensitive training data, embeddings, or real time logs that might tamper with the agent thoughts.\\
\textbf{L3 Agent Frameworks:}The agent planner is qualified prior to running. These are decision path limits, planner validation checks and output limits. These make sure that modules of planning would not swerve into harm or unwanted sequences of tasks.\\
\textbf{L4 Deployment/Infrastructure:}: At the level of deployment and infrastructure, containerization, API rate limits, and zero-trust network policies decrease the attack surface. Both the services are isolated, audited, and bound to the least-privilege permissions.\\
\textbf{L5 Evaluation and observability:}Outlier metrics, behavioral drift and feedback inconsistencies are routinely monitored. These tools provide precursors of insidious exploitation, e.g. slow data poisoning or inferential errors.\\
\textbf{L6 Security \& Compliance:}Regulatory compliance is guaranteed by logging, audit trails, and policy enforcement; root cause analysis is possible. AI explainable modules aid in confirmation of risk-taking decisions.\\
\textbf{L7 Agent Ecosystem:}Trust models are applied to user interfaces, multi-agent gateways and collaborative channels to keep the positive influence during decision-making limited to verified agents or operators.\\
Mapping of specific defense mechanisms to each of MAESTRO layers makes the system tolerant to both horizontal (cross-layer) and vertical (layer-specific) attacks. Suppose one of those layers is compromised, the other layers that are close to each other seal the breach and stop the fix, so that it is a fail-safe dovetailed agent or platform environment.
\section{IMPLEMENTATION AND EVALUATION}
\subsection{SYSTEM SETUP}
The suggested agentic network monitoring system is implemented with the help of the modular-based, event-driven structure that facilitates a real-time analysis and autonomous decision-making. Back end stack is written in Python and uses FastAPI framework to manage asynchronous telemetry ingestion, API routing, and secure communication. This is the infrastructure base on which scalable orchestration services and containerized deployment can be made. The system is used to make sense of traffic intelligently and make informed decisions by combining Pydantic-AI and a large language model (LLM). This architecture allows the agent not only to analyze the traffic characteristics but also to reason the possible anomalies, and make explanations and mitigation suggestions or run them. The LLM works hand in hand with domain-specific prompts and contextual telemetry input to increase the accuracy of agentic actions. 
\begin{figure}[h]
  \centering
  \includegraphics[width=8.5cm,height=5.05cm]{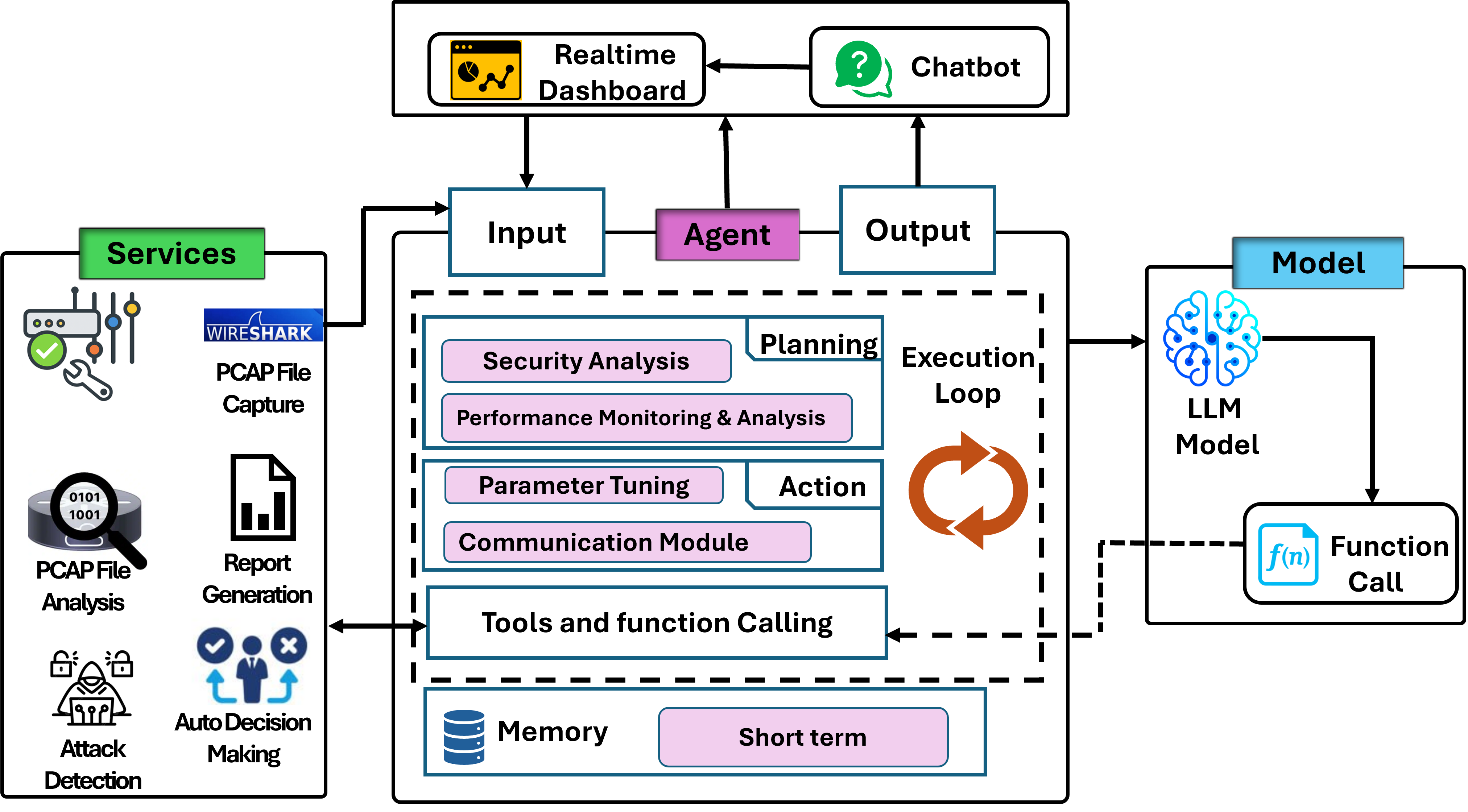}
   \caption{Complete System Architecture}
 %\Description{A woman and a girl in white dresses sit in an open car.}
  \label{fig14}
\end{figure}
To capture network data and parse these, Scapy is used and PCAP files to do traffic simulations during live as well as replay scenarios. At the agent, packet-level features are processed and converted to the structured and consumed telemetry records by the downstream modules. The WebSocket channels are used in a real-time fashion to exchange the telemetry between the backend agent and the frontend interface, their interactions are bi-directional and newly updated messages can be reported and sent on an anomaly and an LLM generated message. The frontend dashboard shows the performance of the system in visual form, active telemetry streams, and interactive reasoning logs of threat flags and processes. It is containerized across the whole system using docker so as to achieve platform independence and maintain reproducibility and also as every component is built within an isolated module so as to accommodate a fault-tolerant system and defense-in-depth approach. The completed architecture is in Figure 5 which presents the layered structure of sensing, analysis, reasoning and human feedback loops.

\subsection{ FUNCTIONAL DEMONSTRATION}
To test the behavior of the agent under malicious conditions such as an adversarial load we did a specific stress test by simulating a resource attack or resource exhaustion form of attack, which fits into Threat 7 (Resource Exhaustion) in the MAESTRO framework. The desired goal was to test the responsiveness, telemetry fidelity and resilience of the system under high network traffic rates and volume similar to the denial-of-service (DoS) condition. tcpreplay was used to replay traffic in a prerecorded PCAP file and was playing on an assigned network interface a traffic originated in a GoldenEye DoS attack. At the same time, performance statistics of the system (both CPU and memory) would be captured in the interval of one second. To simulate a high-load situation, the test was set up to repeat the attack traffic at a rate of 10,000 packets per second in five iterations. Throughout the test, the agent was supposed to keep on checking on traffic in the network and updating its dashboard in real-time.
\begin{figure}[h]
  \centering
  \includegraphics[width=8cm,height=4.57cm]{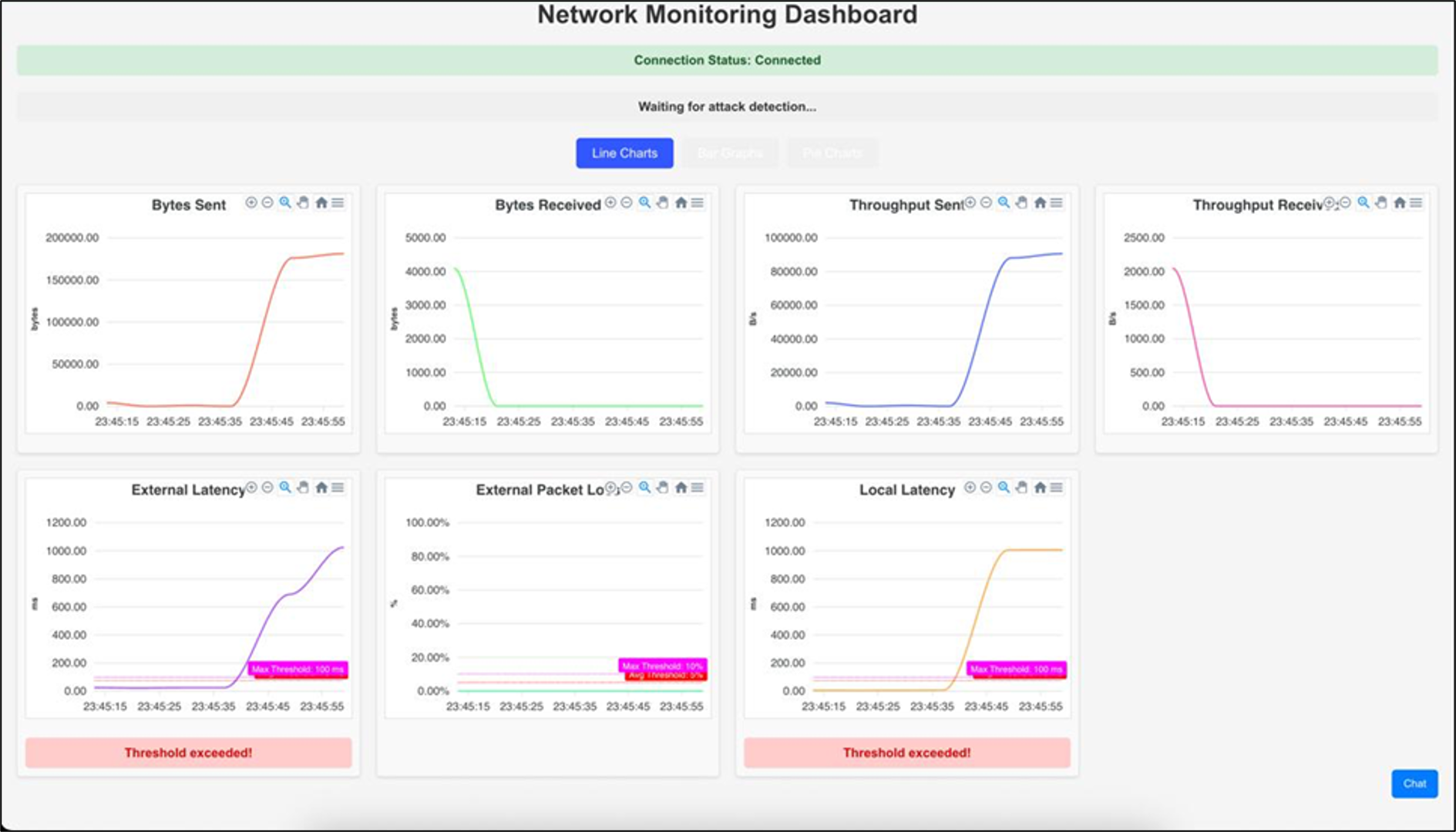}
   \caption{Live Monitoring Dashboard showing metric spikes and threshold violations}
 %\Description{A woman and a girl in white dresses sit in an open car.}
  \label{fig5}
\end{figure}
The telemetry updating occurs every 7-8 seconds on normal operating conditions by the agent. When conducting the high-load test, however, this duration was increased on the scale about 13 times; hence, there was a lag in reporting the metrics and processing the inferences. The system logs and the dashboard response time lag indicated this performance degradation implying that the system was maximum on CPU and memory resources during the simulated environment as shown in Figure 6.

As depicted in Figure 7, these outcomes indicate the susceptibility of the agent to resource-based denial cases and underline the strength of non-static resource monitoring with load cognizant control interventions. The system in the future should include dynamic prioritization of the processes requiring critical telemetry as well as preemptive modules of anomaly mitigation to have consistent velocity of operation under duress.

\begin{figure}[h]
  \centering
  \includegraphics[width=8.5cm,height=4cm]{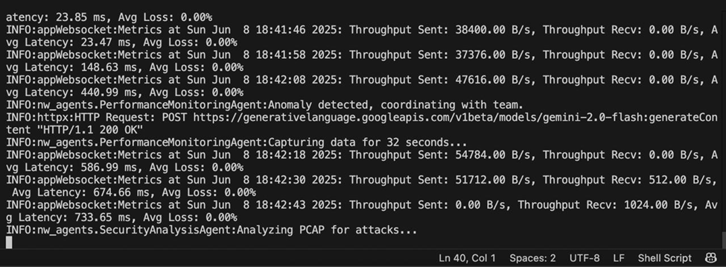}
   \caption{Agent Console Output demonstrating latency analysis, HTTP request tracking, and LLM-based coordination}
   %\Description{A woman and a girl in white dresses sit in an open car.}
  \label{fig6}
\end{figure}
\subsection{SECURITY RISK VALIDATION}
In order to determine the level of robustness of the system against the adversarial influence, two scenarios of high-impact threats were verified against practical test cases related to the MAESTRO threat taxonomy. This segment has the experimental verification of these threats and its influence with respective interpretations.\\
\textbf{Test Case 1 - System Performance when under Network Load (Threat 7: Resource Exhaustion)}\\
\begin{figure}[h]
  \centering
  \includegraphics[width=8.25cm,height=4.02cm]{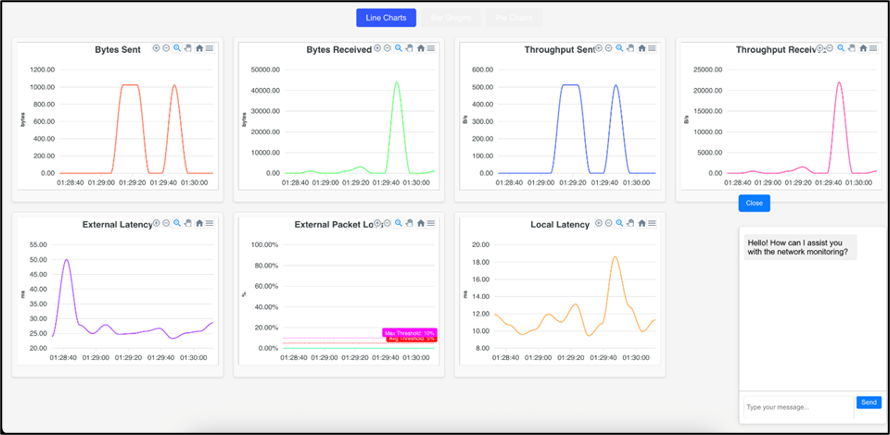}
   \caption{Pre-attack dashboard (before DoS attack)}
   %\Description{A woman and a girl in white dresses sit in an open car.}
  \label{fig7}
\end{figure}
As explained in Section 6.2, a GoldenEye PCAP replay providing a simulation of a DoS attack was applied to the agent. This test provided an important insight that the updating of telemetry is heavily delayed (by 5-6 seconds up to 13 seconds), which indicates the diminishing performance of the agent when used with high resources. The slow reaction time meant that fundamental inductive and making decisions processes were affected once the CPU and memory probability points were reached. This confirms the exploitability of Threat 7 in which the attacker can create havoc to the system in that by consuming excessive computing resources, they can cripple the system without having tampered with the internal system logic.  The test aims at verifying the integrity of the systems in operation under high-load conditions in the modeled way through the replay of the denial-of-service (DoS) attack by the GoldenEye PCAP file. The test was meant to reveal whether the agent was able to retain normal responsiveness when it comes to telemetry amid resource pressure. Under normal conditions, the telemetry dashboard exhibited regular update intervals (approximately every 7–8 seconds), and the system reported stable CPU and memory utilization as shown in Figure 8 .\\
\begin{figure}[h]
  \centering
  \includegraphics[width=8.5cm,height=4.2cm]{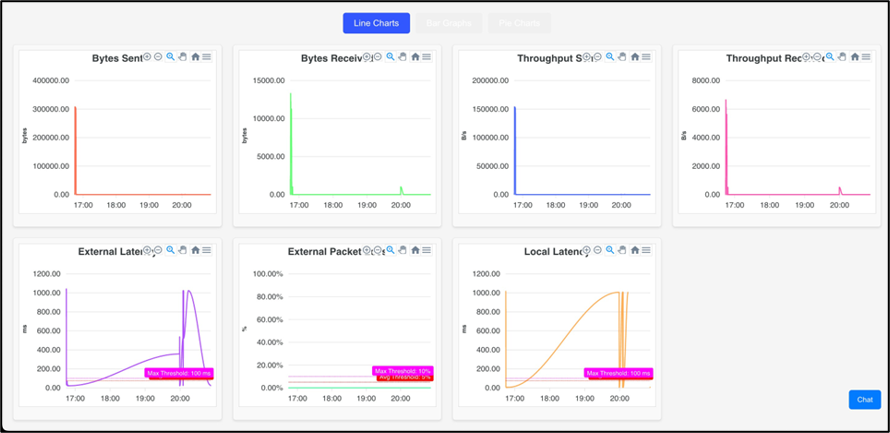}
   \caption{After-attack dashboard (after DoS attack)}
   %\Description{A woman and a girl in white dresses sit in an open car.}
  \label{fig8}
\end{figure}
In the case of PCAP replay, the maximum telemetry update then grew to more than 13 seconds, which shows that the resource was overloaded, which resulted in the performance degradation. The lag polluted the near real-time ability of the agent to analyze and respond as explained in Figure 9.

\textbf{Test Case 2 Memory Poisoning( Threat 8: Knowledge Base Poisoning)}
This test indicates(Figures S10 and S11) how the system performance and behavior is affected by compromised system integrity of memory. The history.json file of the agent that controls the parameter tuning module was manually poisoned and 20 false high-severity attack entries were inserted. Such entries would make the tuning module infer a wrong picture of the threat landscape and would accordingly assign large high durations of packet captures in response to simulated anomalies. The PCAP file created therefore was quite huge compared to the one in the baseline case thus taking higher data processing times by the detection modules and demanding high resources. The memory poisoning damaged efficiency on the detection, but also indirectly led to an exhaustion of the resources, spreading the impact of the Threat 8 to the Threat 7 sequentially. This was demonstrated by attacking a system without changing the logic of the system, but only abusing the memory, which confirms that a low-effort, high-impact attack is possible on unsecured agent memory files. When using valid entries or no recordings, the module chose a reasonable capture time (e.g. 34 seconds) which provided reasonable PCAP sizes and analysis times. 
After the injection of 20 fake high-severity entries, the module modeled high threat environment and correspondingly extended the capture time proportionally to the threat level. The result was a huge PCAP file that slowed down the detecting process and had memory and CPU high use, which is a sign of poor functionality(Figures S12 and S13). \\
Table \ref{tab:risk_validation} summarizes the test cases of security risk validation
\begin{table*}[htbp]
\caption{Summary of Security Risk Validation}
\centering
\small
\setlength{\tabcolsep}{4pt}
\renewcommand{\arraystretch}{0.5}
\begin{tabular}{@{}p{2cm}p{3cm}p{3.8cm}p{3.2cm}p{3cm}p{2cm}@{}}
\toprule
\textbf{Test Case} & \textbf{Threat} & \textbf{MAESTRO Layer(s)} & \textbf{Exploit Method} & \textbf{Observed Impact} & \textbf{Validated Risk} \\
\midrule
TC1: Network Load & Threat \#7: Resource Exhaustion & L4 – Deployment \& Infrastructure, L5 – Evaluation \& Observability & High-speed PCAP replay (DoS) & Delayed telemetry updates; increased CPU/memory usage & Validated \\
\midrule
TC2: Memory Poisoning & Threat \#8: Knowledge Base Poisoning & L2 – Data Operations, L3 – Agent Frameworks & Injected fake history in \texttt{history.json} & Increased capture durations, large PCAP size, detection lag & Validated \\
\bottomrule
\end{tabular}
\label{tab:risk_validation}
\end{table*}
\subsection{LIMITATIONS AND OBSERVATIONS}
 Although the success in the ICMP flood and memory poisoning test cases indicated the promising outcome, there are still a few restrictions in the current version of autonomous network monitoring agent:\\
\textbf{D.1 Single-node Deployment: }The system is currently tested in the single-node environment. This is good to tell the local threat or verify that the inference is correct; and yet this does not help the agent to generalize its multi-agent behavior across distributed or federated environments.\\
\textbf{D.2 Single-node Deployment: }The system is currently tested in the single-node environment. This is good to tell the local threat or verify that the inference is correct; and yet this does not help the agent to generalize its multi-agent behavior across distributed or federated environments.\\
\textbf{D.3 Rule-Based Fall-Backs: }Some of the courses of decision-making in the agent, especially in response time issue and alert escalation remain rather simple sets of rules that are pre-defined than truly autonomous reasoning. This level of automation can limit adaptive behavior in not necessarily simpler or familiar threat environments.\\
\textbf{D.4 Treatment of Subtle Poisoning: }The memory poisoning attack was able to induce a reasoning failure but it can be seen that there is little deeper form of introspection that can occur in LLM-based agents to establish the provenance of input data or mitigate invalid internal lineages of reasoning.\\
\textbf{D.5 Scalability and Resource Load: }Due to the ICMP flooding test, it was identified that the resource handling may become a bottleneck in an event of sudden demand. Subsequent variants will be required to somehow provide dynamic resource scheduling or rate-limiting.\\
\textbf{D.6 Insufficient End-to-End Cryptographic Controls: }Although the telemetry parsing and decision-making modules use access controls, the telecommunication channel in parts of the layers is unencrypted, and it can be a source of risk in the unfriendly network environment.\\
All these findings point out the necessity of more thorough security position and imply distinct paths to follow to improve both architecture and self-awareness potential of the agent.
 
\section{FUTURE DIRECTIONS}
Based on the results of the current research, a number of the directions crucial to promoting agentic AI security using MAESTRO-based modeling is suggested. Such future directions should be used to improve the resilience, scalability and accountability of autonomous systems that are operating dynamically in high-threat environments.\\
\textbf{Multi-Agent Coordination:}Future research could be done in regard to how to demonstrate the structures of safe, collaborative practice of many agents. This involves the design of consensus mechanism, federated reasoning procedure and trust modelling among the agents to avoid intra-agent attacks or domino effect.\\
\textbf{Confidence AI in Adversarial Environments:}It should focus more on adversarial robustness whereby they can develop trust-preserving mechanisms, including the adversarial training process, certifiable reasoning pipelines, and protection against data and model poisoning attacks in all MAESTRO layers.\\
\textbf{Simulated Threat Benchmarks of Agentic AI:}It is necessary to create standardized benchmark suites which would make simulations of the attack situations aiming on definite MAESTRO layers. Reproducibility of these evaluations and the ability to compare these performance in the view of agentic-security frameworks will be made possible as a result of these benchmarks.\\
\textbf{Auditability Policy Compliance and Auditable logging:}Future developments should make sure to work with such frameworks as GDPR, ISO/IEC 27001, and the EU AI Act as the AI systems begin to operate in regulated areas. This also involves introduction of explanable decision logging, real time policy enforcement and secure audit trail.\\
\textbf{Agent Policy Formal Verification:}Vulnerabilities can be prevented by the use of formal methods to prove agent policies and reasoning workflows prior to their deployment. Security invariants can be mathematically checked on all the layers by the aid of model checking, symbolic execution, and SMT solvers.\\
\textbf{Auto-Resilience and Self-Healing Architectures:}Use of self-healing in the agent-architecture will enable autonomous recovery of the agent in the case of a failure or an undeterred attack. Dynamic reconfiguration, hot-patching and anomaly-based containment strategies are encouraging.\\
\textbf{Lightweight and Power-Consuming Security Surveillance:}Future systems especially in edge-based implementations have to be able to balance between energy limitation and real-time monitoring. Such conditions could optimize lightweight cryptographic protocols and adaptive monitoring frequencies and low-power anomaly detectors.\\
\textbf{Cross Layer Security Orchestration:}There must be a centralized orchestration tool that dynamically mappings threat, response, and remediation strategies throughout MAESTRO stack. This would allow smart prioritization and response in accordance to the risk scores and current threat intelligence.\\
    
These guidelines enterprise-wide provide a guide on how to create stronger, more transparent, and scalable systems of agentic AI that can withstand the pressure of adversaries and other such security-related issues.
\section{CONCLUSION}
This study provide a search that ensures a holistic use of the MAESTRO framework to assess and alleviate security threats targeted at agentic AI systems. The layered architecture of the framework allowed a well-organized overview of the threat vectors along the foundational models, data operations, agent frameworks, deployment infrastructure, evaluation modules, security enforcement, and ecosystems. Namely, by putting threats into the context of these seven layers, the study closes the gaps between conceptual taxonomies of threats and implementation vulnerabilities.

The agent system was subjected to the simulation of two critical test scenarios that have been validated to calculate system resilience concerning adversarial conditions. System Performance Under Network Load\_ This was the first scenario which was a denial-of-service attack causing the system to run out of resources. It was found that delayed updates in real-time telemetry occurred, the latency of decisions significantly increased, which indicated the lack of responsiveness during stressful situations. The second scenario, Memory Poisoning, revealed the weaknesses of the parameter tuning module, and the addition of spurious erroneous attack logs into the agent memory (history.json), caused exaggerated capture time, an unreasonably large PCAP file and the decline in performance of the detection pipeline.

The outputs of the two test cases further support the demands of creating agentic AI systems with the subsystems of adaptation, security, and verified integrity. The problems that emerge on the first layer can be carried to the next levels of reasoning, resulting in a chain reaction on the infrastructure and decision results. The cross-layer modeling framework implied by the MAESTRO framework is confirmed by this layered dependency and justified as a framework to realize cross-layer vulnerabilities interconnections.

Along with empirical validations, the study suggested mitigation strategies that would be focused on the specific levels of the MAESTRO levels, such as memory isolation at the data layer, input validation of reasoning modules, sandboxing core APIs and telemetry rollback in response to anomalies. These solutions were graphically depicted and buttressed by a defense-in-depth architecture on how layered resilience may be used to mitigate cross-component threats.

\vfill\pagebreak

\end{document}